\documentclass[12pt]{article}
\usepackage{graphicx}
\usepackage{amssymb}
\begin{document}
\begin{center}
{\Large Isotope effect in yields of nuclear reactions} \\
\vspace{1cm} J.Adam$^{1,4}$, A.R.Balabekyan$^{1,2}$,
A.S.Danagulyan$^{2}$, N.A.Demekhina$^{3}$,
 D.Drnoyan$^{2}$, V.G.Kalinnikov$^{1}$,
G.Musulmanbekov$^{1}$ \\ \vspace{1cm}
\it{1 - Joint Institute for Nuclear Research } \\
\it{2 - Yerevan State University} \\
\it{3 - Yerevan Physics Institute} \\
\it{4 - Institute of Nuclear Physics, R\v{e}z, Czech Republic}
\end{center}

\section{Annotation}

The analysis results of the isotopic yields for products obtained
in reactions of protons at energies 0.66, 1, 8.1 GeV with
separated tin isotopes are presented. The experimental data are
compared in form of reactions of  element  cross sections for two
targets of different isotope compositions. The scaling  behaviour
have been shown in the dependence of this ratios from neutron and
proton composition at the reaction products. The discussion of
scaling parameters allows to explain the mechanism of light and
medium mass fragment production.

\section{Introduction}

Our representation about of mechanism of nuclear interactions
constrained with the characteristics of the emitting products in
the nuclear collision. The complex fragments were generated in the
nuclear reactions of intermediate and hight energies related to
the various processes dynamics of those don't described exactly.

 Recently some scientific nuclear centers have been
carrying out research with ion beams to study the effect of the
isotopic composition of fragments  on their production
probability. Note that despite a lot of measurements of
multiplicities of  light and intermediate-mass fragments and their
characteristics, little is known about the effect of target and
projectile nucleon composition on the formation probability for
various reaction products. Some authors \cite{1}-\cite{9}
investigated ratios between cross sections for formation of light
fragments in targets with different neutron-to-proton composition:

\begin{eqnarray}
R_{21}(N,Z)=Y_{2}(N,Z)/Y_{1}(N,Z)
\end{eqnarray}
where  $Y_{i}(N,Z)$ is the formation cross section for a fragment
with Z protons and N neutrons, $Y_{2}(N,Z)$ is for yields from
targets with a greater number of neutrons than  $Y_{1}(N,Z)$.
Systematization of experimental data \cite{6} according to
isotopic ratios of formation cross sections for light nuclei with
Z = 1-8 made it possible to reveal clear dependence on the nucleon
composition of products and to parameterize relation (1) in the
form of a function of the number of neutrons and protons in the
emitted fragments
\begin{eqnarray}
R_{21}(N,Z) =c \exp (\alpha N +\beta Z)
\end{eqnarray}
where $c$ is the normalizing factor, $\alpha$ and $\beta$ are the
empirical parameters. In \cite{5},\cite{6} the use of three
parameters to describe this dependence for a variety of reactions
was considered as an isotopic scaling.

In the ideal experiment information on primary interaction can be
obtained from the distributions of prefragments resulting from
nuclear collision of about $\sim$ $10^{-21}$ s. Yet, the real time
scale of nuclear fragment measurement is  $\geq 10^{-10}$ s. The
analysis \cite{5},\cite{6} with various model representations
showed that when considering isotopic relation (1), one can ignore
the effect of the slow nucleon emission step for various types of
reaction (multifragmentation, evaporation, deep spallation)
provided that formation of the products in question is of
statistical character and the temperature of excited nuclear
systems is approximately identical.

In the model considered, the coefficients $\alpha$ and $\beta$,
which define factorization of (2), characterize dependence of the
measured distribution on the hot emitter properties, including
proton and neutron density in various transition phases, on
nuclear matter asymmetry allowed for by entering a corresponding
term into the equation of nuclear state, on difference in nucleon
evaporation conditions, etc.

Thus, the method of isotopic relations allows studying the early
reaction step of the decay composite fragmenting system reaction.

 Most of the data analyzed in \cite{5},\cite{6} relate to nucleus-nucleus processes,
which are  characterized by considerable effects resulting from
strong deformation of nuclear matter in collision, such as
compression, rotation, etc. In this sense the study of reaction
induces by light nuclei and particles makes it possible to isolate
the thermal component in nuclear excitation.

According to \cite{7}, the multifragmentation process suggests
decay of a hot nuclear system with formation of light and residual
nuclei with $3 \leq  Z \leq 30$ . Light nuclei can be measured and
identified  and their isotopic properties and formation channels
can be established only with modern detecting facilities including
telescopic systems of high-resolution counters in the 4$\pi$
geometry. Identification of heavy product nuclei is a more
difficult problem. Activation analysis is one of the methods
allowing the nucleon composition of heavy products to be
determined exactly.

The purpose of the presented work was to analyze the ratios of
product yields from reactions induced by 0.66 and 8.1-GeV protons
in targets of separated tin isotopes by the induced activity
method.

Our earlier investigations of the relationship between the
formation cross sections for residual nucleus and the nucleon
composition of the residual and target nuclei revealed correlation
between the yields of these nuclei and their third projection of
the isotopic spin \cite{10}-\cite{12}. Therefore, the isotopic
ratio of the product formation cross sections can be considered in
the form

\begin{eqnarray}
R_{21}(N,Z)=Y_{2}(N,Z)/Y_{1}(N,Z)= \exp(C+Â(N-Z)/2)
\end{eqnarray}

where $C$ and $B$ are the fitting parameters and $(N-Z)/2=t_3$ is
the third projection of the product isotopic spin. This expression
is in general similar to formula (2) in \cite{6}, but the yield
ratios were systematized according to the isospin of the resulting
fragment.

During the element-by-element analysis of formation cross sections
for various isotopic states dependence of the ratio $R_{21}$  on
the number of protons in the product is excluded from (3) and the
neutron content effect defined by the parameter $B/2$ corresponds
to the parameter $\alpha$ in (2) \cite{6}. According to the
theoretical evaluation \cite{1},\cite{13}, \cite{14}, the value of
$\alpha$ depends on a few factors, including the neutron chemical
potential difference $\Delta \mu_n$ in both targets, and on the
value of the asymmetrical energy term appearing in the equation of
nuclear state \cite{5}.

\section{Reactions at 8.1 GeV}

The analysis of experimental data from reactions obtained by
protons at 8.1 GeV energy on different tin isotopes are presented
below. The yields of the reaction products in range $19 \leq Z
\leq 37$ considerably increased at this energy
\cite{10}-\cite{12}, which agrees in general with the predictions
of the statistical multifragmentation model (SMM) \cite{15}. A
distinctive feature of our measurements \cite{16} was a
possibility of investigating dependence of reaction product yields
on the neutron content of the targets used ($^{124}Sn, ^{120}Sn,
^{118}Sn, ^{112}Sn$ ).

We investigated the dependence of $R_{21}(N,Z)$  for the measured
reaction products on the isotopic spin in various target pairs.
Approximating these data in the form of the exponential dependence
(3), we got a set of parameters $B$ characterizing dependence of
the ratio $R_{21}(t3)$ both on the isotopic spin of the product
and on the difference in neutron content of the targets ($\Delta
N=12,8,6,4$). The results of calculations for some reaction
products are presented in Table 1; the dependence in question for
$Br$ isotopes is displayed in Fig. 1. (For clarity, the calculated
curves are shown on the relative scale. The factors introduced for
convenience are given in the figure captions.) It is evident that
within the calculation error all  $R_{21}(t3)$ are characterized
by the identical slope to the $(N - Z)/2$ axis for the targets
with a certain neutron excess and that the dependence curves
become steeper as the difference in the neutron content of targets
increases.

In Fig 2 dependence of the parameter $B$ on the relative neutron
excess of targets is plotted for the target pairs
$^{118}Sn/^{112}Sn, ^{120}Sn/^{118}Sn, ^{124}Sn/^{118}Sn,
^{124}Sn/^{112}Sn$ (solid curve). The relative neutron excess in
targets ($\Delta N$) was used as variable. The parameter is seen
to vary practically linearly with increasing neutron content
difference, the curve slope being 0.11 $\pm$ 0.02.

It is evident from Table 1 that the average value of the parameter
$B$ is 1.1$\pm$0.22 ($\Delta N =12$), which corresponds to
$\alpha$=0.58$\pm$0.06 \cite{5}. This value ($\alpha=0.6$)was used
in \cite{5} to approximate data for evaporation products.

The calculations by the SMM \cite{15} used by us to describe this
type of reactions reflect to an extent the behavior of isospin
characteristics but no full agreement is found on closer
examination (dashed line in Fig. 1).

Cross sections for formation of light fragments in the mass number
region $7\leq A \leq 39$ obtained by us at the proton energy 8.1
GeV \cite{16} were also analyzed by formula (3). As there were not
enough data for the element-by-element analysis, we used all
yields of light products in the above mass region for
approximation. The results of the calculations are presented in
Table 1 and displayed in Fig. 3. The parameter B is seen to vary
with the isotopic spin of both the product and the target. When
isoscaling in reactions of this type is discussed \cite{5},
\cite{6}, the parameter value  $\alpha$=0.37 ($\Delta N=12$) is
used to describe multifragmentation processes. We obtained the
value $B/2=\alpha \sim 0.28\pm0.02$ for  $^{124}Sn/^{112}Sn$ (Fig.
3), which points to multifragmentation rather than evaporation
mechanism for formation of these nuclei at the proton energy 8.1
GeV. Calculations by the SMM \cite{15} showed that the isotopic
dependence in general manifests itself in relative cross sections
for formation of light fragments but particular comparison does
not show satisfactory agreement.

Dependence of the parameter B on the neutron excess in the targets
is shown by a curve with a slope of 0.049$\pm$0.01 (dashed curve
in Fig. 3).

\section{Reactions at 0.66 GeV}

The experimental data in this energy region are not enough to
carry out a thorough element-by-element analysis, as in the case
of 8.1 GeV. Figure 4 (Table 2) present some cross section ratios
from which it is evident that the isospin dependence takes place
at this energy also, including the effect of target neutron
excess. If we take into account closeness of the charge states of
the detected products ($Z = 40-45$) and build an averaged
dependence curve (Fig. 5), all values for  $\Delta N =12$ and
$\Delta N =6$ can be placed within the regions characterized by
the slope parameters $\alpha=0.48\pm0.04$ and
$\alpha=0.295\pm0.015$ respectively. In general, the parameter
values yielded by our analysis agree within the determination
error with the data used in \cite{6} to explain formation of
reaction products through the evaporation. It is seen that for
$\Delta N =12$ the effect of the neutron excess in targets also
manifests itself. The calculations by the cascade model \cite{17}
also yield the isospin dependence of reaction cross sections, but
the values of the coefficients in the approximating functions
turned out to be larger than in the above analysis (dashed line in
Fig. 4).

To get more information on generalities of the isospin-cross
section relationship in nuclear reactions, we carried out a
similar analysis using the experimental data on formation of light
fragments from targets of different isotopic composition in
reactions induced by 0.66 and 1-GeV protons \cite{18}, \cite{19}.
The cited papers point to distinct dependence of the formation
probability for various isotopes of light nuclei on the isotopic
state of target nuclei. Analysis by the scheme proposed in
\cite{6} yielded that  $R_{21}= Y_{2}(t_{3})/Y_{1}(t_{3})$ for the
$H, He, Li, Be, B$, and $C$ produced in $^{124}Sn$ and $^{112}Sn$
targets ($\Delta N=12$) was satisfactorily defined by the
expression $\sim \exp( 0.6 N - 0.82 Z)$ within the 15-20\%
accuracy. The same character of dependence and coefficient values
were taken for description of evaporation processes in \cite{5}.
Thus, we may assume that fragmentation in proton-nucleus
interactions at the energy of 1 GeV and below has predominantly of
the evaporative character.

\section{Conclusion}

Analysis of the experimental data on nuclear reactions in tin
targets of various isotopic composition ($1.24 \leq N/Z \leq 1.48$
) in the region of intermediate and high energies showed distinct
isotopic dependence of product formation cross sections. The
systematics proposed in \cite{5}-\cite{8} allows the assumption
that this dependence is determined by the initial step of the
interaction and the observed isoscaling is due to the statistical
nature of fragment formation. This dependence is reflected in
general in the statistical models used for description of
fragments. Analysis of the coefficients in the expression for the
scaling function on the basis of the data \cite{10}-\cite{12},
\cite{16},\cite{18},\cite{19} showed that yields of light
fragments at the proton energy 8.1 GeV corresponded to the
multifragmentation process ($\alpha=0.28$ ) and at 1 GeV and below
($\alpha=0.6$ ) to the evaporation process. Yields of medium-heavy
and heavy products are due to evaporation and multifragmentation.
Measurements of energy distributions and other experimental data
corroborate the presence of two sources of their formation.
Observation of the dependence in question in a wide range of
energies and reaction types indicates that a universal property of
nuclear interactions manifests itself.

\newpage
\begin{center}
\begin{tabular}{|c|c|c|c|c|c|}  \hline
\label{tabl-1} Residual& \multicolumn{5}{|c|}{Values of the
fitting parameter B($E_{p}=8.1 ÃýÂ$)}
\\ \cline{2-6} nuclei&$\Delta N =4$&$\Delta N =6$&$\Delta N
=8$&$\Delta N =12$& $\alpha_{\Delta N=12}$ \\ \hline
$^{43-47}Sc_{21}$&0.33$\pm$0.13&0.53$\pm$0.08&0.81$\pm$0.16&1.23$\pm$0.20&
0.61$\pm$0.10 \\ \hline
$^{70-74}As_{33}$&0.36$\pm$0.01&0.52$\pm$0.01&-&1.10$\pm$0.22
&0.55$\pm$0.11 \\ \hline
$^{74-77}Br_{35}$&0.34$\pm$0.08&0.63$\pm$0.10&0.81$\pm$0.04&1.16$\pm$0.05&
0.58$\pm$0.05 \\ \hline
$^{76-79}Kr_{36}$&-&-&-&1.10$\pm$0.03&0.55$\pm$0.17 \\ \hline
$^{81-84}Rb_{37}$&0.27$\pm$0.10&0.61$\pm$0.06&0.94$\pm$0.10&
1.23$\pm$0.13&0.61$\pm$0.06 \\ \hline
$^{93-96}Tc_{43}$&0.28$\pm$0.12&-&-&1.22$\pm$0.12&0.61$\pm$0.06 \\
\hline
$^{94-103}Ru_{44}$&-&0.56$\pm$0.07&0.69$\pm$0.04&1.18$\pm$0.08
&0.59$\pm$0.04 \\ \hline $^{103-113}Ag_{47}$&0.37$\pm$0.03&-&-&-&-
\\ \hline the mean&-&-&-&B=1.17$\pm$0.36&$\alpha$=0.58$\pm$0.06 \\
\hline \multicolumn{6}{|c|}{Light fragments ($4 \leq Z \leq 17$)}
\\ \cline{1-6} $\Delta N =2$&$\Delta N =4$&$\Delta N =6$&$\Delta N
=8$&$\Delta N =12$& $\alpha_{\Delta N=12}$ \\ \hline
0.11$\pm$0.02&0.13$\pm$0.05&0.27$\pm$0.10&0.44$\pm$0.05&0.56$\pm$0.02&
0.28$\pm$0.02 \\ \hline
\end{tabular}
\vspace{2cm}

\begin{tabular}{|c|c|c|c|}  \hline
\label{tabl-2} Residual& \multicolumn{3}{|c|}{Values of the
fitting
parameter B ($E_{p}=0.66 ÃýÂ$)} \\
\cline{2-4} nuclei&$\Delta N =4$&$\Delta N =6$&$\Delta N =12$ \\
\hline $^{90-96}Nb_{41}$&0.60$\pm$0.10&-&1.10$\pm$0.14 \\ \hline
$^{91-96}Tc_{43}$&-&0.55$\pm$0.11&- \\ \hline
$^{94-105}Ru_{44}$&-&$0.67\pm$0.05&- \\ \hline
$^{99-105}Ru_{45}$&-&$0.67\pm$0.06&1.30$\pm$0.07 \\ \hline
$^{103-110}Ag_{47}$&-&$0.60\pm$0.12&- \\ \hline
\end{tabular}
\end{center}

\newpage
\begin{figure}[h]
\includegraphics[scale=0.5]{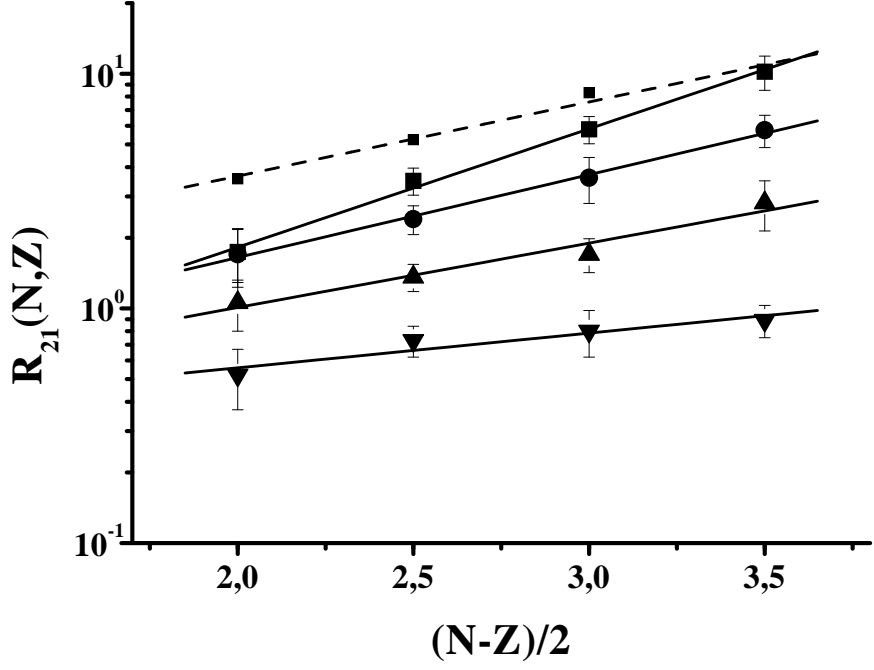}
\caption{ Ratio $R_{21}(t_3)$ versus the third isotopic spin
projection of $^{74-77}Br_{35}$ for various target pairs (Ep = 8.1
GeV). Experimental points:  $\blacksquare - -\Delta N = 12,
(\times 10)$; $\bullet - \Delta N = 8, (\times 5)$;
$\blacktriangle - \Delta N = 6, (\times 2)$; $\blacktriangledown -
\Delta N = 4, (\times 1)$. Solid lines are the results of fitting
by (3), the dashed line is the calculation by the SMM for $\Delta
N = 12, (\times 10)$ [33].}
\end{figure}
\newpage
\begin{figure}[h]
\includegraphics[scale=0.5]{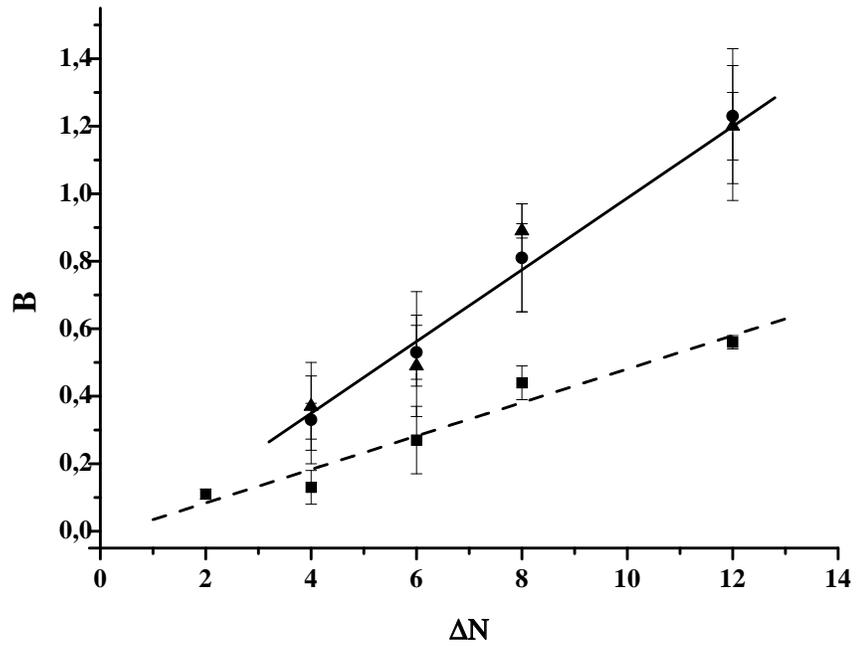}
\caption{ Parameter $B$ versus the difference in the neutron
number ($\Delta N$) of targets (Ep = 8.1 GeV). The solid curve is
for medium heavy residual nuclei, the dashed curve is for light
ones.}
\end{figure}
\newpage
\begin{figure}[h]
\includegraphics[scale=0.5]{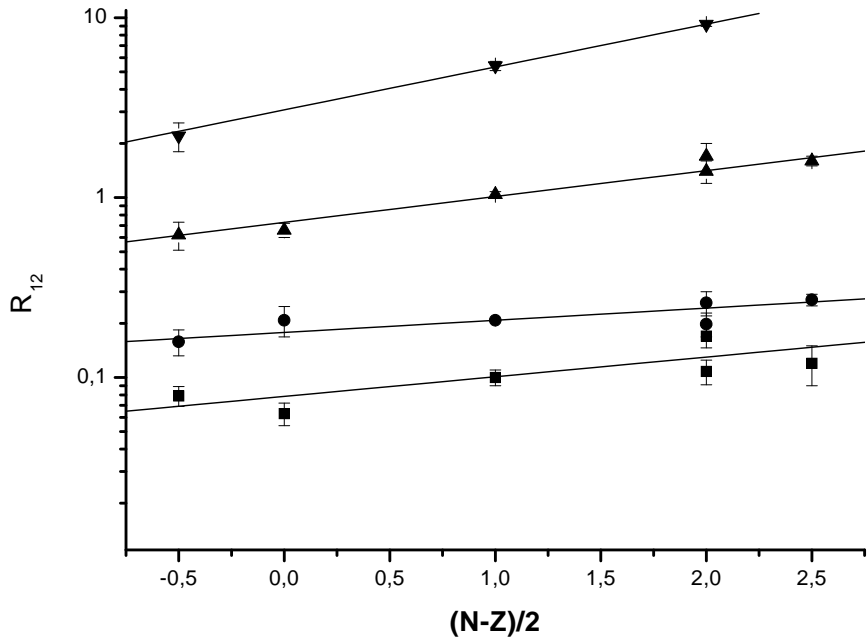}
\caption{ Ratio $R_{21}(t_3)$ versus the third isotopic spin
projection of light fragments ( $^{7}Br, ^{22,24}Na, ^{28}Mg,
^{38,39}Cl$) for various target pairs (Ep = 8.1 GeV). Experimental
points:  $\blacktriangledown - \Delta N = 12, (\times 4);
\blacktriangle - \Delta N = 6, (\times 1); \bullet - \Delta N = 4,
(\times 0.2); \blacksquare - \Delta N =2, (\times 0.1)$. Solid
lines are the results of fitting by (3).}
\end{figure}
\newpage
\begin{figure}[h]
\includegraphics[scale=0.5]{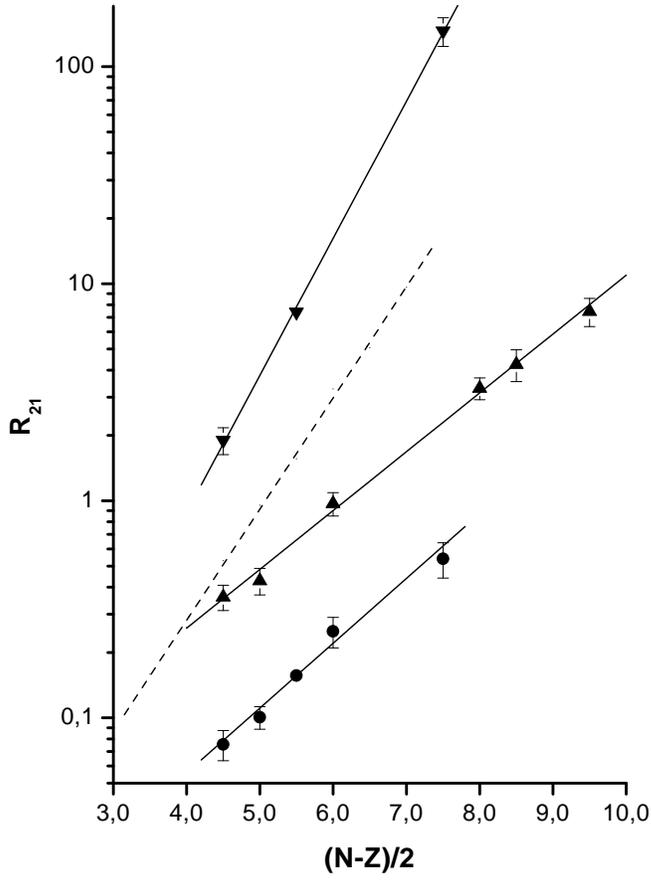}
\caption{ Ratio $R_{21}(t_3)$ versus the third isotopic spin
projection for various target pairs (Ep = 0.66 GeV). Experimental
points:  $\bullet - ^{99-105}Rh, \Delta N = 6, (\times 0.2);
\blacktriangle - ^{103-113}Ag, \Delta N = 6, (\times 1);
\blacktriangledown - {99,101,105}Rh, \Delta N = 12, (\times 10)$.
Solid lines are the results of fitting by (3), the dashed line is
the calculation for $Rh, \Delta N = 6, (\times 1)$ [16].}
\end{figure}
\newpage
\begin{center}
\begin{figure}[h]
\includegraphics[scale=0.5]{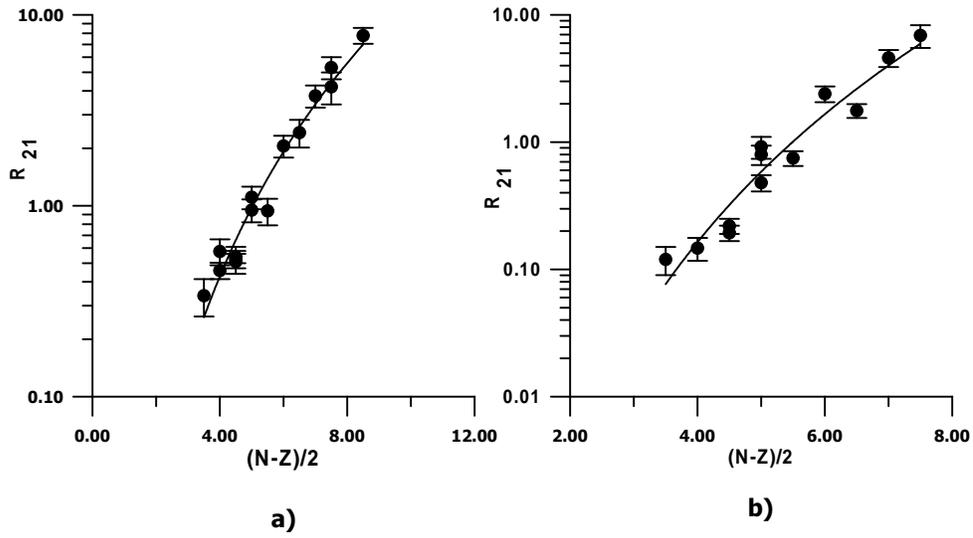}
\caption{ Ratio $R_{21}(t_3)$ versus the third isotopic spin
projection of products in the charge region Z = 40-45 (Ep = 0.66
GeV). $\bullet$ - experimental points, solid lines are the results
of fitting by (3). (a) $\Delta N = 6$, (b) $\Delta N = 12$.}
\end{figure}
\end{center}
\end{document}